\documentclass[amsmath,amssymb,aps, onecolumn,
preprint,superscriptaddress,
]{revtex4-2}

\begin{document}

\title{Bosonisation as the Hubbard Stratonovich Transformation}

\author{I. V. Yurkevich}

\address{School of Physics \& Astronomy, The University of Birmingham}

\date{\today}
\maketitle

\section{Introduction }

The problem of strongly correlated electrons in one dimension attracted the attention
of condensed matter physicists since early 50's. After the seminal paper of Tomonaga \cite{Tom:50}
who suggested the first soluble model in 1950, there were essential achievements reflected in
papers by Luttinger \cite{Lutt:63} (1963) and Mattis and Lieb \cite{ML:65} (1963). Considerable 
contribution to understanding of the generic properties of the 1D electron liquid has been done by 
Dzyaloshinskii and Larkin \cite{DzLar:73} (1973) and Efetov and Larkin \cite{EfLar:76} (1976). 
Despite the fact that the main
features of the 1D electron liquid were captured and described by the end of 70's, the
investigators felt dissatisfied with the rigour of the theoretical description.
The most famous example is the paper by Haldane \cite{Hal:81} (1981) where the author developed the
fundamentals of the modern bosonisation technique,  known as the operator approach.
This paper became famous because the author has rigourously shown how to construct the Fermi 
creation/anihilation operators out of Bose ones. The most recent example of dissatisfaction
is the review due to von Delft and Schoeller \cite{vDS:98} (1998) who revised the approach
to the bosonisation and came up with what they called constructive bosonisation.

There exists another way to bosonise the problem. It was suggested 25 years ago by \cite{Fogedby:76}
and found its further elaboration in the paper by Lee and Chen \cite{LeeChen:88} in 1988.
This approach (sometimes called "functional bosonisation") can be 
found in variety of papers (see, for example, \cite{Naon} and \cite{Kop}). 
The main objective of this lecture is to assemble these pieces in a
clear picture. The aim is purely methodological which is the reason why
the number of complications (like backscattering, presence of lattice and spin degrees
of freedom) are kept at the lowest possible level. Any generalisations can be easily 
incorporated into the scheme. The final formulation of the bosonised problem will look
distinct from what can be found elsewhere. We believe that the approach present in this
lecture is the "shortest shortcut" from the original fermionic problem of interacting electrons
to the problem of non-interacting bosons.

\section{Functional Integral Formulation}

We start from a general Hamiltonian describing interacting electrons in one dimension:

\begin{equation}
{\hat{\cal H}}=\frac{1}{2m}\int{\rm d}x\partial_x{\hat\Psi}^{\dagger}(x)\partial_x{\hat\Psi}(x) \\
+\frac{1}{2}\int{\rm d}x{\rm d}x'{\hat\Psi}^{\dagger}(x){\hat\Psi}^{\dagger}(x')\,V_0(x-x')\,
{\hat\Psi}(x'){\hat\Psi}(x).
\end{equation}
Here $V_0(x-x')$ is the bare inter-electron interaction which is usually taken to be the Coulomb
interaction screened by the carriers of the media surrounding the one-dimensional system. 

The thermal Green function is the time-ordered product of the Fermion anihilation and 
creation operators 

\begin{equation}
{\cal G}(x,\tau;x',\tau')=\left\langle T_{\tau}
{\hat\Psi}(x,\tau){\hat{\bar\Psi}}(x',\tau')\right\rangle
\end{equation}
defined in the imaginary time Heisenberg representation
\begin{equation}
\nonumber
{\hat\Psi}(x,\tau)
=e^{{\hat{\cal H}}'\tau}{\hat\Psi}(x)e^{-{\hat{\cal H}}'\tau},\quad
{\hat{\bar\Psi}}(x,\tau)
=e^{{\hat{\cal H}}'\tau}{\hat\Psi}^{\dagger}(x)e^{-{\hat{\cal H}}'\tau},
\end{equation}
where ${\hat{\cal H}}'={\hat{\cal H}}-\mu{\hat N}$. Here $\mu$ is the chemical potential
and ${\hat N}$ is the particle number operator. This Green function
can be written in the form of the coherent state functional integral
over the Grassmann fields that reflects the fermionic nature of electrons:

\begin{equation}
\label{F}
{\cal G}(x,\tau;x',\tau')=\frac{\int{\cal D}\psi{\cal D}\psi^*\psi(x,\tau)\psi^*(x',\tau')
\exp\left\{-{\cal S}[\psi,\psi^*]\right\}}
{\int{\cal D}\psi{\cal D}\exp\left\{-{\cal S}[\psi,\psi^*]\right\}},
\end{equation}
where the action ${\cal S}[\psi,\psi^*]$ is defined by

\begin{equation}
{\cal S}[\psi,\psi^*]=\int{\rm d}x{\rm d}\tau\psi^*(x,\tau)\left[\partial_{\tau}+
{\hat\xi}\right]\psi(x,\tau) 
+\frac{1}{2}\int{\rm d}x{\rm d}x'{\rm d}\tau\psi^*(x,\tau)\psi^*(x',\tau)\,V_0(x-x')\,
\psi(x',\tau)\psi(x,\tau).
\end{equation}
Here  ${\hat\xi}=-(2m)^{-1}\partial^2_x-\mu$ is the kinetic energy operator counted from 
the chemical potential $\mu$, and
$\psi(x,\tau)$ is the Grassmann function (or, equivalently, fermionic field) which is an
anti-periodic function of the imaginary time $\tau$, $\psi(\tau+\beta)=-\psi(\tau)$,
with the inverse temperature $\beta=T^{-1}$ being the period.

Being aware that individual electrons are not good as a starting point for description
of excitations in one dimension we choose to introduce an auxiliary field conjugated to
the electron density. This can be achieved via the Hubbard-Stratonovich transformation which
brings the fermionic problem to a Gaussian form at the price of complication caused by
appearance of some new field and related to it extra integration. Tthis is similar to the introduction of an 'order parameter field' in superconducting systems \cite{nlsm} with the essential difference that the mean classical value of the field is zero. We start from the following 
representation of the exponential containing the interaction

\begin{equation}
\label{HS}
\exp\left[-{\cal S}_{int}\right]
=\frac{\int{\cal D}\phi\exp\left[-\frac{1}{2}\phi V_0^{-1}\phi+i\phi \psi^*\psi\right]}
{\int{\cal D}\phi\exp\left[-\frac{1}{2}\phi V_0^{-1}\phi\right]},
\end{equation}
where we used symbolic expressions
\begin{eqnarray}
\nonumber
\phi V_0^{-1}\phi & \equiv &\int{\rm d}x{\rm d}x'{\rm d}\tau\phi(x,\tau)V_0^{-1}(x-x')\phi(x',\tau),\\
\nonumber
\phi \psi^*\psi & \equiv & \int{\rm d}x{\rm d}\tau\phi(x,\tau)\psi^*(x,\tau)\psi(x,\tau)
\end{eqnarray}
The notation $V_0^{-1}(x-x')$ stands for the inverse function understood as
an integral kernel inverse to that corresponding to the bare interaction:

\begin{equation}
\int{\rm d}x_1V_0^{-1}(x-x_1)V_0(x_1-x')=\delta(x-x').
\end{equation}
One can see that the field $\phi(x,\tau)$ should posess the same properties as a product of
two Grassmann fields $\psi^*(x,\tau)\psi(x,\tau)$, i.e. it is a commuting field and periodic
under the shift $\tau\to\tau+\beta$. Therefore, this field is of bosonic nature.
Now we substitute the Hubbard-Stratonovich transform, Eq.(\ref {HS}), into the original representation
Eq.(\ref {F}) of the thermal fermionic Green function:

\begin{equation}
\label{mix}
{\cal G}(x,\tau;x',\tau')=\frac{\int{\cal D}\psi{\cal D}\psi^*{\cal D}\phi\,
\psi(x,\tau)\psi^*(x',\tau')
\exp\left\{-{\cal S}[\psi,\psi^*;\phi]\right\}}
{\int{\cal D}\psi{\cal D}\psi^*{\cal D}\phi\exp\left\{-{\cal S}[\psi,\psi^*;\phi]\right\}}.
\end{equation}
Now the action is a functional of not only the fermionic field but also of the auxiliary field
$\phi$:
\begin{equation}
\label{Gau}
{\cal S}[\psi,\psi^*;\phi]
=\frac{1}{2}\int{\rm d}x{\rm d}x'{\rm d}\tau
\phi(x,\tau)V_0^{-1}(x-x')\phi(x',\tau)
+\int{\rm d}x{\rm d}\tau\psi^*(x,\tau)\left[\partial_{\tau}+
{\hat\xi}-i\phi(x,\tau)\right]\psi(x,\tau)
\end{equation}
Now we divide and multiply both numerator and denominator by

\begin{equation}
\int{\cal D}\psi{\cal D}\psi^*
\exp\left\{-\int{\rm d}x{\rm d}\tau\psi^*\left[\partial_{\tau}+
{\hat\xi}-i\phi\right]\psi\right\}={\rm Det}\left[\partial_{\tau}+{\hat\xi}-i\phi\right]
=\exp\left\{{\rm Tr}\ln\left[\partial_{\tau}+{\hat\xi}-i\phi\right]\right\}.
\end{equation}
Then one can see that the thermal Green function can be written as some auxiliary Green
function ${\tilde{\cal G}}[\phi]$

\begin{equation}
\label{GF}
{\tilde{\cal G}}(x,\tau; x',\tau';[\phi])
=\frac{\int{\cal D}\psi{\cal D}\psi^*  \psi(x,\tau)\psi^*(x',\tau')
\exp\left\{-\int{\rm d}x{\rm d}\tau\psi^*\left[\partial_{\tau}+{\hat\xi}-i\phi\right]\psi
\right\}}
{\int{\cal D}\psi{\cal D}\psi^*\exp\left\{-\int{\rm d}x{\rm d}\tau\psi^*
\left[\partial_{\tau}+{\hat\xi}-i\phi\right]\psi\right\}}
\end{equation}
averaged over the field $\phi$ with the new bosonic action ${\cal S}[\phi]$:

\begin{equation}
\label{bos}
{\cal G}(x,\tau; x',\tau')
={\cal Z}^{-1}\int{\cal D}\phi\,{\tilde{\cal G}}(x,\tau; x',\tau';[\phi])\,e^{-{\cal S}[\phi]}
\end{equation}
where
\begin{equation}
\label{bos1}
{\cal Z}=\int{\cal D}\,\phi\,e^{-{\cal S}[\phi]}, \qquad
{\cal S}[\phi]
=\frac{1}{2}\phi V_0^{-1}\phi
-{\rm Tr}\ln\left[\partial_{\tau}+{\hat\xi}-i\phi\right].
\end{equation}

The auxiliary Green function is the one of the "free" electrons in the spatial and temporal dependent
field $\phi(x,\tau)$ and it must be found from the equation

\begin{equation}
\label{tGF}
\left(\partial_{\tau}+{\hat\xi}-i\phi(x,\tau)\right){\tilde{\cal G}}(x,\tau; x',\tau';[\phi])
=\delta(x-x')\,\delta(\tau-\tau').
\end{equation}
If we were able to solve this equation for arbitrary field $\phi(x,\tau)$ we would then have
some expression for ${\tilde{\cal G}}$ which depends only on the bosonic field $\phi(x,\tau)$.
Substituting this solution of Eq.(\ref{tGF}) into Eq.(\ref{bos}) we would finally reduce our 
problem to another one where we must do some integral over the bosonic field governed by the action
Eq.(\ref{bos1}). And it should be pointed out that this procedure can be implemented for any
observables constructed from any number of Fermi operators, not only for a bilinear combination
defining the one-particle Green function Eq.(\ref{F}). The only extra effort we would need is to use
the Wick theorem in Eq.(\ref{GF}) which is valid because the action is quadratic in 
$\psi$-fields after the Hubbard-Stratonovich transformation is implemented.

It is evident enough, that this scheme is absolutely general and therefore absolutely useless.
We should refer to some specifics of one-dimensional problems to develop it to the stage where
it becomes a working tool. And that is what we will do next.

\section{One Dimension}

The main feature of one dimensional problems with Fermions is that their Fermi
surface consists of just two points $-p_F$ and $+p_F$. Accordingly, electrons close
to this Fermi surface can be divided into two groups, left- (with the momenta in the
vicinity of $-p_F$) and right-movers (with the momenta in the vicinity of $-p_F$).
That means that the Fermi fields over which we integrate in Eq.(\ref{F}) might be
represented in the form

\begin{equation}
\psi(x,\tau)=\psi_+(x,\tau)\,e^{ip_Fx}+\psi_-(x,\tau)\,e^{-ip_Fx},
\end{equation}
We are interested in large scale behaviour of our system which makes this distinction
well defined because only small vicinity of the Fermi points should be taken into account.
In other words, the Fermi fields $\psi_{\pm}(x,\tau)$ may be considered as smooth functions
on the scale $p_F^{-1}$. It allows as when writing action in Eq.(\ref{F}) to act by the kinetic
energy operator ${\hat\xi}$ on the fast oscillating exponential mainly, neglecting second order
derivatives of $\psi_{\pm}(x,\tau)$. Also, the integral over the whole space suppresses 
contribution from the "uncompensated" oscillatory functions proportional to $e^{\pm i2p_Fx}$.
Therefore, we may introduce two-vector

\begin{equation}
\Psi^T(x,\tau)=(\psi_+(x,\tau),\psi_-(x,\tau))
\end{equation}
and in these notations the action Eq.(\ref{bos1}) becomes
\begin{equation}
\label{22}
{\cal S}[\phi]=\frac{1}{2}\phi\,V_0^{-1}\,\phi
+{\rm Tr}\ln\left(
\begin{array}{cc} 
\partial_{\tau}-iv_F\partial_{x}-i\phi & 0 \\
0 & \partial_{\tau}+iv_F\partial_{x}-i\phi
\end{array}\right),
\end{equation}
Writing down this action we neglected the backscattering of electrons due to Coulomb interaction.
Generally speaking in the Coulomb field $\phi(x,\tau)$ there are two essential components
related to two different scattering events. One of them contains the Fourier harmonics with
$|q|\ll p_F$ and, therefore, is a smooth function of $x$, another one consists of the Fourier
harmonics with $|q|\approx 2p_F$. The former is responsible for the forward scattering, i.e.
when two colliding electrons keep their direction of motion which they had before collision.
The latter causes backscattering when two colliding electrons make the U-turn after the
collision. If we assume that for the Fourier transforms of the bare interaction we have
the inequality $V_0(q\ll 2p_F)\ll V_0(q\approx 2p_F)$, we may neglect the backscattering which
means that the Coulomb field $\phi(x,\tau)$ in the action Eq.(\ref{22}) is a smooth function
on the scale of the electron wavelength $p_F^{-1}$. 

The Green function now becomes a $2\times 2$ matrix and neglecting backscattering we find that
it is a diagonal matrix with the diagonal elements ${\tilde{\cal G}}_+$ and ${\tilde{\cal G}}_-$
defined by the equations:
\begin{equation}
\label{tgf}
\left(\partial_{\tau}\mp iv_F\partial_{x}-i\phi(x,\tau)\right)
{\tilde{\cal G}}_{\pm}(x,\tau; x',\tau';[\phi])
=\delta(x-x')\,\delta(\tau-\tau').
\end{equation}
The formal solution of these equations can be easily written

\begin{eqnarray}
\label{Gtilde}
{\tilde{\cal G}}_{+}(x,\tau; x',\tau';[\phi])=g_{+}(x-x',\tau-\tau')\,
e^{i\theta(x,\tau)-i\theta(x',\tau')}; \\
\nonumber
{\tilde{\cal G}}_{-}(x,\tau; x',\tau';[\phi])=g_{-}(x-x',\tau-\tau')\,
e^{i\theta^*(x,\tau)-i\theta^*(x',\tau')},
\end{eqnarray}
where $\theta(x,\tau)$ should be determined from another equation

\begin{equation}
\label{theta}
\left(\partial_{\tau}- iv_F\partial_{x}\right)\theta(x,\tau)=\phi(x,\tau)
\end{equation}
and $g_{\pm}=\left(\partial_{\tau}\pm iv_F\partial_{x}\right)^{-1}$ are the Green functions 
of the chiral particles:

\begin{equation}
g_{\pm}(x,\tau)=\frac{T}{2v_F}\sin^{-1}\pi T\left(\tau\mp i\frac{x}{v_F}\right).
\end{equation}

Now, when we have conveniently expressed the auxiliary Green function as an exponential
of the linear functional of the $\phi$-field, one should take a closer look at the action
itself. The term written as ${\rm Tr}\ln$ in Eq.(\ref{22}) is a sum of two terms which are the
contributions of the right- and left-movers. Let us analyze the right-movers contribution first.
We may expand ${\rm Tr}\ln$ in powers of $\phi$:

\begin{equation}
{\rm Tr}\ln\left(\partial_{\tau}- iv_F\partial_{x}-i\phi(x,\tau)\right)
={\rm Tr}\ln\left(\partial_{\tau}- iv_F\partial_{x}\right)
-\sum_{n=1}^{\infty}\frac{1}{n}{\rm Tr}\left(g_+\,i\phi\right)^n.
\end{equation}
The term containing the $n$-th power of $\phi$ is the vertex containing the loop of $n$-th
order $\Gamma_n^+$ and $n$ external lines corresponding to $\phi$'s:

\begin{equation}
\label{loop}
{\rm Tr}\left(g_+\,\phi\right)^n=\int\prod_{k=1}^{n}{\rm d}x_k{\rm d}{\tau}_k\;
\Gamma_n^+(x_1,\tau_1;...;x_n,\tau_n)\,\prod_{i=1}^{n}\phi(x_i,\tau_i).
\end{equation}
The loop is built from chiral Green functions $g_+$ in the following way:

\begin{equation}
\Gamma_n^+(x_1,\tau_1;...;x_n,\tau_n)=\prod_{i=1}^{n}g_+(x_i-x_{i+1},\tau_i-\tau_{i+1})
\end{equation}
and $(x_{n+1},\tau_{n+1})=(x_1,\tau_1)$. It is convenient to calculate this loop in
the frequency/momentum representation. We use notations $\epsilon$ and $\Omega$ for the
fermionic and bosonic Matsubara frequencies correspondingly. Due to translational invariance
of the system the loop $\Gamma_n^+$ will depend on $n-1$ momenta and frequencies

\begin{equation}
\Gamma_n^+(q_1,\Omega_1;...;q_n,\Omega_n)
=T\sum_{\epsilon}\int\frac{{\rm d}p}{2\pi}\;g_+(p,\epsilon)g_+(p+q_1,\epsilon+\Omega_1) 
\cdot ...\cdot  g_+(p+q_1+...+q_{n-1},\epsilon+\Omega_1+...+\Omega_{n-1}).
\end{equation}
Using explicit expression of the Green function of right-moving free electrons
\begin{equation}
g_+(p,\epsilon)=\left(-i\epsilon+v_Fq\right)^{-1}
\end{equation}
and introducing temporarily the notations

\begin{equation}
z=-i\epsilon+v_Fq, \qquad z_i=-i\Omega_i+v_Fq_i
\end{equation}
we can write the expression for the $n$-th loop in the following form

\begin{equation}
\Gamma_n^+
=\int{\rm d}z\;\frac{1}{z\,(z+z_1)\,(z+z_1+z_2)...(z+z_1+z_2+...+z_{n-1})},
\end{equation}
where we have used a symbolic notation for
$\int{\rm d}z = T\sum_{\epsilon}\int\frac{{\rm d}p}{2\pi}$. There are very widely used
in QED the Feynman identities which help calculating higher order diagrams. One of them,

\begin{equation}
\frac{1}{a_1...a_2}=\int\limits_{0\leq x_{n-1}\leq ...\leq x_1\leq 1}
\frac{(n-1)!\,{\rm d}x_1 ...{\rm d}x_{n-1}}
{\left[a_1x_{n-1}+a_2(x_{n-2}-x_{n-1})+...+a_n(1-x_1)\right]^n}
\end{equation}
after some modification, may be brought to the form which is of help in our case, namely

\begin{equation}
\label{Fe}
\Gamma^+_n(z_1,...,z_{n-1})=(n-1)!\int{\rm d}z\;\int\limits_{Y_{n-1}}
{\rm d}^{n-1}y\left[z+\sum_{i=1}^{n-1}y_i\,z_i\right]^{-n}.
\end{equation}
where the inner integral runs over the area $Y_{n-1}=(0\leq y_{n-1}\leq ...\leq y_1\leq 1)$.
All loops $\Gamma^+_n$ appear only in integrals with the products of $\phi$'s (see Eq.(\ref {loop})) 
which are explicitely symmetric with respect to the permutation of all arguments $(x_i,\tau_i)$
(or, equivalently, with respect to the permutations of $(q_i,\Omega_i)$ in frequency/momentum
representation).
Therefore, only symmetrized form $\Gamma^{+\, ({\rm sym})}_n$ contributes to the integral in 
Eq.(\ref {loop}). Symmetrization of of the expression (\ref{Fe}) leads to discarding the factorial
and relaxing the ordering of $y$'s:

\begin{equation}
\label{F1}
\Gamma^{+\, ({\rm sym})}_n(z_1,...,z_{n-1})=\int{\rm d}z\;\int\limits_0^1
{\rm d}^{n-1}y\left[z+\sum_{i=1}^{n-1}y_i\,z_i\right]^{-n}.
\end{equation}
For $n > 2$ these integrals vanish. To see that we may take one integral over some $y_i$, 
say $y_{n-1}$:
\begin{equation}
\Gamma^{+\, ({\rm sym})}_n\propto\int{\rm d}z\int\limits_0^1
{\rm d}^{n-2}y\left[\left(z+z_{n-1}+\sum\limits_{i=1}^{n-2}y_i\,z_i\right)^{-n+1}
-\left(z+\sum\limits_{i=1}^{n-2}y_i\,z_i\right)^{-n+1}\right].
\end{equation}
For $n>2$ the integral of each term in the square brackets is finite. It means that we can make a shift 
of the integration variables in each term independently. Making the shift $z+z_{n-1}\to z$ 
(meaning $p+q_{n-1}\to p, \, \epsilon+\Omega_{n-1}\to \epsilon$) in the first term (integral) we
find that it becomes identical to the second one and, thus, cancel it.
Therefore, all loops containing more than two external lines are zero.
That was the observation first made by Dzyaloshinskii and Larkin \cite{DzLar:73}. Above I gave the
simplest proof to that fact. Nowdays this cancellation when it happens in many different
one-dimensional problems is called the Loop Cancellation theorem. 

Therefore, we are left with only two contributions from the loops with one and two legs.
Calculating them one must excercise precaution because the corresponding integrals are ill-defined.
These troubles have their roots in the linearization of the spectrum. However, we may
predict the correct result even without healing these problems. It is clear that the loop
with one leg is proportional to the zero mode of the Coulomb interaction and unperturbed
particle density. This term is always cancelled against positive background. The loop with two
legs is the standard polarization operator which is know to be (for the right-movers) in
$(q,\Omega)$-representation

\begin{equation}
\pi_+(q,\Omega)=\frac{1}{2\pi}\frac{q}{-i\Omega+v_Fq}
\end{equation}
Finally, back to the action Eq.(\ref{22}), we collect two contributions coming from both
chiral species and notice that the part independent of $\phi$ drops out of the expression
Eq.(\ref{tGF}). The action then becomes

\begin{equation}
{\cal S}[\phi]=\frac{1}{2}\phi V_0^{-1}\phi+\frac{1}{2}\phi \pi\phi,
\end{equation}
where we introduced the overall polarization operator which is the sum of two chiral
polarization operators $\pi_+$ and $\pi_-$ and has the following familiar form in
$(q,\Omega)$-representation

\begin{equation}
\pi(q,\Omega)=\pi_{+}(q,\Omega)+\pi_{-}(q,\Omega)
=\nu\frac{v^2_Fq^2}{\Omega^2+v^2_Fq^2},
\end{equation}
with $\nu=(\pi v_F)^{-1}$ being the one-dimensional density of states. 

The Gaussian
action describing Coulomb bosons $\phi$ can be cast in a familiar form if we introduce
effective time/frequency-dependent interaction

\begin{equation}
\label{SC}
{\cal S}[\phi]=\frac{1}{2}\,\phi\,V^{-1}\,\phi,
\end{equation}
where

\begin{equation}
V^{-1}=V_0^{-1}+\pi, \qquad V^{-1}(q,\Omega)=V_0^{-1}(q)+\pi(q,\Omega)
\end{equation}
One can easily recognize the well-known result for the effective Coulomb potential with
the screening calculated in the random phase approximation.
It means that the random phase approximation becomes exact as we linearize the spectrum. 

\section{Calculation of Observables}

After the Hubbard-Stratonovich transformation is done the action Eq.(\ref{Gau})
becomes quadratic in the fermionic fields and Wick's theorem becomes applicable
for any product of the $\psi$-fields. Consequently, any observable, originally
a product of $\psi$-operators, decomposes into the product of the auxiliary Green functions
${\tilde{\cal G}}_{\pm}[\phi]$ which should be averaged over the auxiliary field $\phi$.
This field describes non-interacting bosons whose propagator is the screened Coulomb
interaction Eq.(\ref{SC}). The auxilary Green functions ${\tilde{\cal G}}_{\pm}[\phi]$ are products
of the free electron Green function $g_{\pm}$ and the exponentials with the exponent
being a linear functional of the Coulomb bosons $\phi$ (see Eqs.(\ref{Gtilde})and (\ref{theta})). 
Therefore, the problem of interacting 
electrons is reduced to the calculation of the simple Gaussian integrals.

In this way we, as an example, may calculate the Green function of the right movers
\begin{equation}
\left\langle{\tilde{\cal G}}_+(x,\tau;[\phi])\right\rangle_{\phi}=g_+(x,\tau) ,
\exp\left[-{\cal K}(x,\tau)\right]
\end{equation}
where the exponent is given by
\begin{equation}
{\cal K}(x,\tau)=\frac{1}{2}\langle\left[\theta(x,\tau)-\theta(0,0)\right]^2\rangle 
=T\sum_{\Omega}\int\frac{{\rm d}q}{2\pi}\frac{V(q,\Omega)}{(-i\Omega+v_Fq)^2}
\left[e^{iqx-i\Omega\tau}-1\right].
\end{equation}
The propagator of the "screened Coulomb field" ($\phi$-field) is
\begin{equation}
V(q,\Omega)=V_0(q)\frac{\Omega^2+v_F^2q^2}{\Omega^2+v^2q^2},
\end{equation}
where the renormalized velocity $v$ is defined as
\begin{equation}
v(q)=v_F\sqrt{1+\frac{V_0(q)}{\pi v_F}}.
\end{equation}
The simple arithmetics gives us the result which can be found in any review on one-dimensional
bosonisation:
\begin{equation}
{\cal K}(x,\tau)=\int\frac{{\rm d}q}{q}\left\{\left[
\frac{e^{-v_Fq\tau}}{e^{-\beta v_Fq} + 1}-\frac{e^{-vq\tau}}{e^{-\beta vq} + 1}
\right]e^{iqx}\right. 
+\left.\frac{(v-v_F)^2}{2vv_F}\frac{1-\cos qx e^{-vq\tau}}
{e^{-\beta vq} + 1}
\right\}.
\end{equation}
It is not our goal to describe the behaviour of different response functions. The results
obtained with the use of the formalism present in this lecture are exactly the same as those
obtained within operator formalism due to Haldane. The main distinction of the approach
given in this lecture is its simplicity.

\section{Electrons with Spin}
We accept "$g$-ology" as our starting point, neglecting Umklapp and backscattering:
\begin{eqnarray}
\displaystyle
S[\psi]=\int{\rm d}x{\rm d}\tau
\psi^{*}_{\sigma\alpha}(x,\tau)\left[\partial_{\tau}
-ir_{\alpha}v_F\partial_{x}\right] \psi_{\sigma\alpha}(x,\tau) \\
+ \frac{1}{2}\int{\rm d} x{\rm d} x'{\rm d}\tau
g_{\sigma\sigma'}^{\alpha\alpha'}(x-x')
n_{\sigma\alpha}(x,\tau)n_{\sigma'\alpha'}(x',\tau),
\end{eqnarray}
where $\sigma=\uparrow,\downarrow$ describes spin dependence and $\alpha=\pm$
refers to right- ($+$) or left-moving ($-$) particles ($r_{\pm}=\pm 1$). The coupling 
matrix has the structure
\begin{equation}
g_{\sigma\sigma'}^{\alpha\alpha'}
=\delta_{\alpha,\alpha'}\left[\delta_{\sigma,\sigma'}g_4^{\parallel}
+\delta_{\sigma,-\sigma'}g_4^{\perp}\right]
+\delta_{\alpha,-\alpha'}
\left[\delta_{\sigma,\sigma'}g_2^{\parallel}
+\delta_{\sigma,-\sigma'}g_2^{\perp}\right].
\end{equation}
After Hubbard-Stratonovich transformation we arrive at
\begin{equation}
S[\phi]=\frac{1}{2}\phi_s g^{-1}_{ss'} \phi_{s'} 
- \sum_{s}{\rm Tr}\ln
\left[\partial_{\tau}-ir_{\alpha}v_F\partial_{x}-i\phi_{s}\right]
\end{equation}
Please note that we combine spin and chirality in one composite index $s=(\sigma,\alpha)$, 
where possible. Then expanding $'{\rm Tr\ln}'$ as in the previous section
we are again left with the RPA result:
\begin{equation}
\label{RPAs}
S[\phi]=S_{\bf RPA}[\phi]=\frac{1}{2}\phi_s V^{-1}_{ss'}\phi_{s'}, \quad
V^{-1}_{ss'}=g^{-1}_{ss'}+\delta_{s,s'}\pi_{\alpha}
\end{equation}
The matrix Green function is diagonal in the spin indices and is the sum of two chiral
functions:
\begin{equation}
{\cal G}_{\sigma\sigma'}(x-x',\tau-\tau')=
\sum_{\alpha,\alpha'}\langle\psi_{\sigma\alpha}(x,\tau)\psi^*_{\sigma'\alpha'}(x',\tau')\rangle
e^{ip_F(r_{\alpha}x-r_{\alpha'}x')} 
=\delta_{\sigma\sigma'}\sum_{\alpha}{\cal G}_{\sigma\alpha}(x-x',\tau-\tau')
e^{ip_Fr_{\alpha}(x-x')},
\end{equation}
where
\begin{equation}
{\cal G}_{\sigma\alpha}(x-x',\tau-\tau')
= g_{\alpha}(x-x',\tau-\tau')
\left\langle 
e^{i[\theta_{s}(x,\tau)-\theta_{s}(x',\tau')]}
\right\rangle.
\end{equation}
The brackets stand for the averaging over $\theta_s$ with the action Eq.(\ref{RPAs}).
The field $\theta_{s}$ itself is a linear functional of the "Coulomb field" $\phi_s$, and must be
found as the solution to the following equation
\begin{equation}
\left(\partial_{\tau}-ir_{\alpha}v_F\partial_{x}\right)\theta_{s}({\bf x})
=\phi_{s}({\bf x}).
\end{equation}
Taking the Gaussian integral over $\phi$ we obtain
\begin{equation}
\displaystyle
\left\langle 
e^{i[\theta_s(x,\tau)-\theta_{s}(x',\tau')]}
\right\rangle
=e^{-{\cal K}_s(x-x',\tau-\tau')}, 
\end{equation}
with the exponent related to the diagonal matrix element of the screened 'Coulomb interaction'
$V_{ss}(q,\Omega)$ (see Eq.(\ref{RPAs})):
\begin{equation}
{\cal K}_s(x,\tau)
=T\sum_{\Omega}\int\frac{{\rm d}q}{2\pi}\frac{V_{ss}(q,\Omega)}{(-i\Omega+v_Fq)^2}
\left[e^{iqx-i\Omega\tau}-1\right] 
\end{equation}

Due to the fact that the free Hamiltonian (and thus, density-density correlators
$\pi_{\pm}({\bf q})$) do not depend on spin, we can make a rotation in the spin
sub-space only to diagonalize the coupling matrix $g_{ss'}$. The rotation can be performed
by the matrix which has the form 
\begin{equation}
L=\frac{1}{\sqrt 2}\left(\begin{array}{cc}1 & 1 \\[4pt]
1 & -1 \end{array}\right)
\end{equation}
in the representation where the four-vector composed of components 
$\phi_{\sigma\alpha}$ is taken to be $\phi^{\rm T}=(\phi_{\uparrow +},\phi_{\uparrow -},
\phi_{\downarrow +}, \phi_{\downarrow -})$. This rotation, in fact, is equivalent to saying
that we are writing the quadratic form in the new fields $\phi_{\rho\alpha}$ and $\phi_{\nu\alpha}$
which are the linear combinations of the old ones:
\begin{equation}
\phi_{\rho\alpha}=\frac{1}{\sqrt 2}(\phi_{\uparrow\alpha} +\phi_{\downarrow\alpha}),
\quad
\phi_{\sigma\alpha}=\frac{1}{\sqrt 2}(\phi_{\uparrow\alpha} -\phi_{\downarrow\alpha}),
\end{equation}
corresponding to charge- (index $\rho$) and spin- (index $\sigma$) fields. After
such a rotation the action becomes
\begin{equation}
S[\phi]=\frac{1}{2}\sum_{\nu,\alpha,\alpha'}\phi_{\nu\alpha} 
V^{-1}_{\nu,\alpha\alpha'}\phi_{\nu\alpha'},
\end{equation}
where $\nu=\rho,\sigma$. The new interaction matrix is block-diagonal in its spin/charge sectors
and each block can be represented as a 2-by-2 matrix in the chiral indices:

\begin{equation}
{\hat V}_{\nu}=\frac{1}
{(a_{4\nu}+\pi_+)(a_{4\nu}+\pi_-)-a^2_{2\nu}}
\left(
\begin{array}{cc}
\displaystyle
a_{4\nu}+\pi_-  & a_{2\nu}\\
\displaystyle
a_{2\nu} & a_{4\nu}+\pi_+
\end{array}\right)
\end{equation}
where we introduced spin and charge coupling constants
\begin{eqnarray}
g_{i\rho}=\frac{1}{2}(g_i^{\parallel}+g_i^{\perp}), \quad
g_{i\sigma}=\frac{1}{2}(g_i^{\parallel}-g_i^{\perp}), \quad
a_{i\nu}=\frac{g_{i\nu}}{2(g^2_{4\nu}-g^2_{2\nu})}.
\end{eqnarray}
Substituting explicit expressions for the chiral polarization operators $\pi_{\pm}(q,\Omega)$
one can write it in the following form:
\begin{equation}
{\hat V}_{\nu}(q,\Omega)=2\pi\,
\frac{\Omega^2+v_F^2q^2}{\Omega^2+v_{\nu}^2q^2}\,{\hat A}(q,\Omega)
\end{equation}
where ${\hat A}(q,\Omega)$ is 2-by-2 matrix
$$
{\hat A}(q,\Omega)=\left(
\begin{array}{cc}
\displaystyle(v-v_{\nu})-2\gamma_{\nu}v_{\nu}\frac{-i\Omega+v_Fq}
{i\Omega+v_Fq} & 
-\frac{v_{\nu}}{2}(K_{\nu}-K_{\nu}^{-1}) \\
 -\frac{v_{\nu}}{2}(K_{\nu}-K_{\nu}^{-1})& \displaystyle
(v-v_{\nu})-2\gamma_{\nu}v_{\nu}\frac{i\Omega+v_Fq}
{-i\Omega+v_Fq}
\end{array}\right)
$$
written in the notations standard for the literature on the Luttinger Liquid:

\begin{eqnarray}
v_{\nu}^{\pm}=v_F+\frac{1}{\pi}(g_{4\nu}\pm g_{2\nu}),\quad
K_{\nu}=\sqrt{\frac{v_{\nu}^{-}}{v_{\nu}^{+}}}, \\
\nonumber
v_{\nu}=\sqrt{v_{\nu}^{+}v_{\nu}^{-}},\quad
\gamma_{\nu}=\frac{1}{4}\left(K_{\nu}+K^{-1}_{\nu}-2\right)
\end{eqnarray}

The diagonal part of the interaction matrix $V_{ss}$ needed for the average Green function
does not actually depend on spin index $\sigma$:
\begin{equation}
V_{ss}=\frac{1}{2}\sum_{\nu}V_{\nu,\alpha\alpha}
\end{equation}
Therefore, for the average Green function we have the representation: 
\begin{equation}
{\cal G}(x,\tau)=\sum_{\alpha}g_{\alpha}(x,\tau)\,e^{-{\cal K}_{\alpha}(x,\tau)}
e^{ip_Fr_{\alpha}x}, 
\qquad
{\cal K}_{+}(x,\tau)={\cal K}_{-}(-x,-\tau)
\end{equation}
where
\begin{equation}
{\cal K}_{+}(x,\tau)=\sum_{\nu}\int\frac{{\rm d}q}{q}\left\{\frac{1}{2}\left[
\frac{1-e^{-v_{\nu}q\tau+iqx}}{e^{-\beta v_{\nu}q}+1}
-\frac{1-e^{-v_Fq\tau+iqx}}{e^{-\beta v_Fq}+1}
\right]\right. 
\left.-\gamma_{\nu}
\frac{e^{-v_{\nu}q\tau}}{e^{-\beta v_{\nu}q}+1}\cos{qx}\right\}
\end{equation}

Once again, the expression above is well known and can be found in any review on the Luttinger
Liquid. The task of this Lecture was to show that starting from the field-theoretical description
of one-dimensional interacting electrons 
( by this I mean representation of the observables as functional integral over the Grassmann fields )
one can derive the free-bosons action in extremely simple, almost trivial, and elegant way.


\end{document}